\documentclass{article}
\usepackage{color}
\usepackage{graphicx}
\usepackage{amsmath, amsthm}
\usepackage{mathtools}
\usepackage{natbib}
\usepackage{url}
\RequirePackage[colorlinks,citecolor=blue,urlcolor=blue]{hyperref}
\usepackage{xcolor}
\usepackage{chngcntr}
\usepackage{subfig}
\usepackage{hyperref}
\usepackage{algorithm}
\usepackage[noend]{algpseudocode}
\usepackage{setspace}

\usepackage{xfrac}
\definecolor{forest}{rgb}{0.133,0.545,0.133}
\usepackage{multirow}
\usepackage{amsfonts}

\usepackage{enumitem}
\usepackage{caption}
\usepackage{array}
\usepackage{makecell}

\evensidemargin=\oddsidemargin
\usepackage{etoolbox}

\newif\ifabbreviation
\pretocmd{\thebibliography}{\abbreviationfalse}{}{}
\AtBeginDocument{\abbreviationtrue}

\begin{document}
	\newcommand{\bb}{\boldsymbol{\beta}}

	\title{Bayesian Sample Size Determination: Sampling Distribution Estimation or Exploration?}


	\author{Luke Hagar\footnote{Luke Hagar is the corresponding author and may be contacted at \url{luke.hagar@ubc.ca}.} \hspace{35pt} 
    Paul Gustafson 
    \\ 
 \textit{Department of Statistics, The University of British Columbia}}


	\date{}

	\maketitle

	\begin{abstract}

To design many Bayesian studies, the sample size is chosen to attain sufficient power to reject a null hypothesis or a desired probability that an interval estimate is sufficiently narrow. Determining the minimum suitable sample size for studies with complex models typically requires intensive Monte Carlo simulation. One simulation-based strategy to consider the sample-size space involves estimating sampling distributions of posterior summaries, such as posterior probabilities or interval estimate lengths, at various sample sizes. Another strategy instead explores these sampling distributions via stochastic root finding over the sample-size space, where the entire sampling distribution of posterior summaries is not estimated at any given sample size. The performance of these competing approaches can be assessed using their Monte Carlo sampling distributions of sample size recommendations across repeated implementations. In this paper, we theoretically derive these Monte Carlo sampling distributions for various Bayesian sample size determination approaches. To promote best practices for Bayesian sample size determination, we conduct extensive numerical studies to estimate these Monte Carlo sampling distributions for the sampling distribution estimation and exploration approaches to assess their bias, precision, and sensitivity to tuning parameters.

		\bigskip

	\end{abstract}

	\maketitle

	\baselineskip=19.5pt



\section{Introduction}\label{sec:intro}

To ensure the reliability and precision of scientific findings, sample size determination (SSD) is a critical component of study design. In recent decades, Bayesian methods for data-driven decision making have become increasingly popular in scientific studies. Many methods for Bayesian SSD exist, including decision theoretic and performance-based approaches. Decision theoretic approaches directly incorporate utility functions and select the sample size that maximizes expected utility (see e.g., \citet{raiffa1961applied,lindley1997choice}). Performance-based approaches \citep{spiegelhalter1994bayesian, brutti2014bayesian} do not directly incorporate utility functions and instead aim to control inference within a specified degree of error. Performance-based approaches are common in the design of Bayesian experiments, including clinical trials and reliability studies (see e.g., \citet{wang2002simulation, berry2010bayesian, stevens2022cpm, golchi2024estimating}). 

This paper focuses on performance-based approaches defined via power or precision criteria. These criteria are defined for a target of inference $\theta$ with respect to various assumptions about the data-generating process. In this paper, power criteria for study design are used to select the minimum sample size $n$ that attains desired power to reject a null hypothesis about $\theta$ using posterior probabilities. The precision criteria in this paper are defined with respect to a length criterion \citep{joseph1997bayesian, santis2004two} that is satisfied when the credible interval (CI) for $\theta$ is sufficiently narrow. The recommended sample size is the minimum $n$ for which the length criterion is satisfied with desired probability.

Simulation is typically required to find minimum sample sizes that satisfy performance criteria for Bayesian studies with complex statistical models. \citet{wang2002simulation} proposed a general framework for Bayesian SSD based on estimating sampling distributions of posterior summaries via Monte Carlo simulation. In this framework, a sample size $n$ is selected. Many samples of that size are simulated from a prior predictive distribution \citep{gubbiotti2011bayesian, brutti2008robust} that incorporates assumptions about the data-generating process. The relevant posterior summary (i.e., posterior probability or CI length) is recorded for each sample, and the collection of posterior summaries estimates their sampling distribution. From this sampling distribution estimate, power or the probability of satisfying the length criterion can be approximated for this value of $n$. 

That computationally intensive process is repeated for various $n$ until a suitable sample size is found, and a strategy must be selected to decide which values of $n$ to assess via simulation. In simulation-based SSD more broadly, grid searches \citep{wang2002simulation,landau2013sample,green2016simr} and bisection methods \citep{stevens2022cpm, shilane2023simulation} are often recommended to search the $n$-space. While less common, simulation-based SSD approaches that estimate sampling distributions for consecutively increasing $n$ until the performance criteria are satisfied have been proposed (see e.g., \citet{jiang2012practical}). These strategies to explore the $n$-space do not account for Monte Carlo error in the estimates of power or the probability of satisfying the length criterion, which can introduce bias into the sample size recommendation. They also ignore information from large-sample theory \citep{vaart1998bvm} pertaining to sampling and posterior distributions, thereby inefficiently searching the $n$-space. Nevertheless, these commonly adopted strategies serve as benchmarks for more sophisticated SSD methods.

By leveraging large-sample theory, Bayesian SSD based on power \citep{hagar2025economical} and precision \citep{hagar2025precision} criteria can be implemented by estimating the sampling distribution of posterior summaries at only two values of $n$. Given these two sampling distribution estimates, a set of linear proxy functions can be fit to estimate power or the probability of satisfying the length criterion throughout the $n$-space. Although the use of these proxies to efficiently search the $n$-space is justified using asymptotic theory \citep{vaart1998bvm}, good performance  of these linear proxy functions in Bayesian SSD has been observed for finite $n$ in many settings \citep{hagar2025economical,hagar2025precision}. However, the bias and variance associated with the resulting sample size recommendations have not been formally studied.

Instead of estimating the entire sampling distribution of posterior summaries at various $n$ values, Bayesian SSD can be implemented using stochastic root finding. One popular and straightforward approach is the Robbins-Monro (RM) algorithm \citep{robbins1951stochastic}, a stochastic analog to Newton's method that is appropriate when the underlying function (e.g., the power curve) can only be approximated with noisy estimates. In each iteration of RM, the posterior summary for a single simulated sample of size $n$ is recorded before selecting the sample size for the next iteration. Because only a single posterior summary is obtained in each iteration, their sampling distribution is \emph{explored} but not estimated at any $n$ value. 

RM and its extensions have been incorporated into various SSD methods \citep{amaratunga1999searching, sommer2026carts, sadatsafavi2026bayesian, hagar2024fast}, but the approach was initially proposed for univariate functions with a continuous input. In that setting, RM theoretically converges to the root given various assumptions about the underlying function and step sizes between successive iterations; however, convergence issues have been well documented in practice \citep{nemirovski2009robust, toulis2021proximal}. RM and its extensions may facilitate SSD by rounding the value of $n$ to the nearest integer in each iteration. Yet, the impact of this rounding on the RM algorithm's convergence to the correct sample size recommendation has not formally been studied. 

In this work, we compare Bayesian SSD carried out using the fundamentally different strategies of estimating and exploring sampling distributions. The SSD framework with linear proxy functions and the RM algorithm serve as efficient exemplars of the two strategies. While more complex proxy functions \citep{wilson2021efficient,golchi2024estimating} or stochastic optimization methods \citep{valko2013stochastic, grill2015black} could serve as exemplars, this additional complexity is not needed when the unknown function is a monotonic and smooth function of $n$ (e.g., a power curve). 

We formally compare these two approaches by considering Monte Carlo sampling distributions (MCSDs) of sample size recommendations across repeated implementations. We formally prove that MCSDs for the approach with linear proxy functions are approximately normal. This new theoretical result complements a similar result for the RM algorithm in continuous root-finding settings \citep{fabian1968asymptotic, polyak1992acceleration}. It is thus natural to compare MCSDs for the two approaches using mean squared error with respect to true sample size recommendations. To inform best practices for Bayesian SSD, these comparisons consider the robustness of the MCSDs to various tuning parameters.

 The remainder of this article is structured as follows. In Section \ref{sec:prelim}, we introduce preliminary concepts for the Bayesian SSD methods compared in this paper along with the power and precision criteria that these methods are based on. In Section \ref{sec:mcsd}, we formally prove that MCSDs for the SSD approach with linear proxy functions are approximately normal and describe the MCSDs derived for the remaining methods. We conduct numerical studies to compare the performance of the SSD approaches based on estimating and exploring sampling distributions and make resulting recommendations in Section \ref{sec:num}. We conclude with a summary and discussion of extensions to this work in Section \ref{sec:disc}.

\section{Preliminaries}\label{sec:prelim}

\subsection{Performance Criteria}\label{sec:prelim.perf}

We begin by defining the performance-based design criteria considered in this paper. Such criteria aim for pre-experimental probabilistic control over inference about a target of interest $\theta$. In pre-experimental settings, we represent data from a future, random sample of size $n$ as $\boldsymbol{Y}_n$. The observed data are denoted by $\boldsymbol{y}_n = \{y_i\}_{i=1}^n$. We suppose that the target of inference is a function of the parameters $\boldsymbol{\eta}$ for the statistical model: $\theta = g(\boldsymbol{\eta})$. 
 
 A \emph{design} prior $p_D(\boldsymbol{\eta})$ \citep{de2007using,berry2010bayesian,gubbiotti2011bayesian} incorporates uncertainty regarding the parameters in pre-experimental settings. The design prior is typically informative and concentrated on parameter values that are relevant to the study objectives, so it may differ from the \emph{analysis} prior used to analyze the observed data. The design prior defines the prior predictive distribution of $\boldsymbol{Y}_n$:
\begin{equation}\label{eq:prior_pred}
p(\boldsymbol{y}_n) = \int \prod_{i = 1}^{ n } f(y_{i}; \boldsymbol{\eta}) \hspace{1pt} p_D(\boldsymbol{\eta}) \hspace{1pt} d\boldsymbol{\eta},
\end{equation}
where $f(y; \boldsymbol{\eta})$ is the model for the data. The model $f(\cdot)$ may also depend on additional covariates $\boldsymbol{x}$, but we do not formally incorporate those covariates into the posterior predictive distribution for simplicity.

The distribution in (\ref{eq:prior_pred}) may be defined under the conditional or predictive approach \citep{gubbiotti2011bayesian}. The conditional approach uses a degenerate design prior 
$p_D(\boldsymbol{\eta})$ such that the model parameters take a value of $\boldsymbol{\eta}_{0}$ for all hypothetical repetitions of an experiment. The predictive approach instead uses a nondegenerate $p_D(\boldsymbol{\eta})$. For each hypothetical repetition of the experiment, different parameter values $\boldsymbol{\eta} \sim p_D(\boldsymbol{\eta})$ characterize the data-generating process. This paper broadly accommodates both the conditional and predictive approaches. 

The performance criteria for power and precision are defined when the data are generated via (\ref{eq:prior_pred}).  Power-based SSD is appropriate when the goal of the experiment is to conclude that $H_1 : \theta \in (\delta_L, \delta_U)$, where $-\infty \le \delta_L < \delta_U \le \infty$ defines some general interval hypothesis based on superiority, non-inferiority, or practical equivalence. We select $n$ to ensure the probability of correctly concluding that $H_1 : \theta \in (\delta_L, \delta_U)$ is true is at least $1 - \beta$ for some target power $1 - \beta \in (0, 1)$. For analyses facilitated via posterior probabilities, the chosen sample size ensures that
\begin{equation}\label{eq:power_prob}
\Pr\{\Pr(H_1 ~| ~ \boldsymbol{Y}_n) \ge \gamma\} \ge 1 - \beta,
\end{equation}
for some decision threshold $\gamma \in [0.5,1)$. To \emph{correctly} conclude that $H_1$ is true in (\ref{eq:power_prob}), we must select a design prior $p_{D}(\boldsymbol{\eta})$ such that $p_{D}(H_1) = 1$. 

Precision-based SSD is appropriate when the study goal is to precisely estimate $\theta$ using a two-sided CI with credibility level $1 - \alpha \in (0,1)$. The length criterion \citep{joseph1997bayesian,santis2004two} is satisfied when the length of this CI for $\theta$ does not exceed a target length $l > 0$. We select $n$ to ensure the probability of satisfying the length criterion is at least $q$:
\begin{equation}\label{eq:prec_prob}
\Pr\{L_{1 - \alpha}(\boldsymbol{Y}_n) \le l\} \ge q,
\end{equation}
where $L_{1 - \alpha}(\boldsymbol{Y}_n)$ is the length of the $100 \times (1-\alpha)\%$ CI and $q \in (0,1)$ is a target probability. Unlike with power-based SSD, there is no restriction on the support of $p_{D}(\boldsymbol{\eta})$ when the goal of the experiment is related to precision. In the subsequent subsections, we overview several SSD approaches that can be used to find the minimum $n$ such the criterion in (\ref{eq:power_prob}) or (\ref{eq:prec_prob}) is satisfied.
 
\subsection{Design with Linear Proxy Functions}\label{sec:prelim.lin}

We next overview the main SSD framework we consider that is based on sampling distribution estimation. This framework based on linear proxy functions requires that we estimate the sampling distribution of posterior summaries at only two sample sizes: $n_0$ and $n_1$. We first introduce this framework in the context of power-based SSD. As $n \rightarrow \infty$, \citet{hagar2025economical} proved that the quantiles of the sampling distribution of the logit of posterior probabilities are linear functions of $n$ under the \emph{conditional} approach, where the limiting slope is a function of the effect size. That limiting result holds true when the regularity conditions for (i) the asymptotic normality of the maximum likelihood estimator (MLE) for $\boldsymbol{\eta}$ and (ii) the Bernstein-von Mises (BvM) theorem are satisfied. The MLE conditions in Theorem 5.39 of \citet{vaart1998bvm} are incorporated into the BvM conditions in Theorem 10.1 of \citet{vaart1998bvm}. The conditions for the BvM theorem additionally require that the prior distribution for $\boldsymbol{\eta}$ is continuous with positive density at the true parameter value $\boldsymbol{\eta}_0$. When $\theta = g(\boldsymbol{\eta})$ is a continuous function of the model parameters, the induced prior for $\theta$ also satisfies these conditions.

To adapt those asymptotic trends for SSD, \citet{hagar2025economical} obtained sampling distribution estimates at $n_0$ and $n_1$ via simulation: $\{\Pr\left(H_1 ~| ~ \boldsymbol{y}_{n_0, r}\right)\}_{r=1}^m$ and $\{\Pr\left(H_1 ~| ~ \boldsymbol{y}_{n_1, r}\right)\}_{r=1}^m$, where $\boldsymbol{y}_{n, r}$ denotes the sample of size $n$ simulated from (\ref{eq:prior_pred}) in Monte Carlo iteration $r = 1, \dots, m$. They then used these two sampling distribution estimates to fit linear functions of $n$ whose $y$-values are the $r^{\text{th}}$ order statistics of $\left\{\text{logit}\left(\Pr\left(H_1 ~| ~ \boldsymbol{y}_{n_0, s}\right)\right)\right\}_{s=1}^m$ and $\left\{\text{logit}\left(\Pr\left(H_1 ~| ~ \boldsymbol{y}_{n_1, s}\right)\right)\right\}_{s=1}^m$ at $x = n_0$ and $n_1$. Across $r = 1, \dots, m$, those empirically fit functions served as linear proxy functions to model the sampling distribution of posterior probabilities throughout the $n$-space. Power at a new sample size $n$ can quickly be estimated as the proportion of the $m$ linear functions whose $y$-values at $x=n$ are at least $\text{logit}(\gamma)$. The recommended sample size is the minimum $n$ for which this power estimate based on the linear proxy functions is at least $1 - \beta$. 

To extend SSD to the predictive approach, the logits at $n_0$ and $n_1$ were split into $B$ bins based on the effect size corresponding to $\boldsymbol{\eta} \sim p_{D}(\boldsymbol{\eta})$ in each Monte Carlo iteration $r$ before fitting the linear functions. A non-degenerate design prior $p_D(\boldsymbol{\eta})$ marginalizes the sampling distribution of posterior probabilities with respect to different effect sizes for $\theta$. Because the slope of the linear functions depends on the effect size, this binning procedure roughly discretizes the marginal sampling distribution to reduce bias when applying theoretical results pertaining to the conditional approach where the effect size is fixed. A formal algorithm for power-based SSD with linear proxy functions is reproduced in Appendix A.1 of the supplement.

 We next discuss this framework in the context of precision-based SSD. As $n \rightarrow \infty$, \citet{hagar2025precision} proved that the median of the sampling distribution of the logarithm of CI lengths is a linear function of $\log(n)$ under the conditional approach. Furthermore, they showed that the logarithm of the absolute difference between other sampling distribution quantiles and the median is asymptotically linear with respect to $\log(n)$. Both results again hold true under the conditions for the asymptotic normality of the MLE and those for the BvM theorem.

To implement SSD, sampling distribution estimates $\{L_{1-\alpha}( \boldsymbol{y}_{n_0, r})\}_{r=1}^m$ and $\{L_{1-\alpha}( \boldsymbol{y}_{n_1, r})\}_{r=1}^m$ were obtained. The medians of $\{\log(L_{1-\alpha}( \boldsymbol{y}_{n_1, r}))\}_{r=1}^m$ and $\{\log(L_{1-\alpha}( \boldsymbol{y}_{n_0, r}))\}_{r=1}^m$ were extracted to construct a linear function of $\log(n)$. The logarithm of the absolute differences between $\log(L_{1-\alpha}( \boldsymbol{y}_{n, r}))$ and those medians were then computed for each sample size. The order statistics of these logarithmic differences at $n_0$ and $n_1$ were used to fit a second set of linear functions with respect to $\log(n)$. The two sets of functions served as linear proxy functions to model the sampling distribution of CI lengths across the $n$-space. The probability of satisfying the length criterion can quickly be estimated; the recommended sample size is the minimum $n$ for which this estimated probability based on the linear proxy functions is at least $q$. Binning was again used to accommodate precision-based SSD under the predictive approach as detailed in the algorithm from \citet{hagar2025precision} that we reproduce in Appendix A.2.

The tuning parameters for this SSD framework consist of the sample sizes $n_0$ and $n_1$. We note that the choice of the second sample size $n_1$ can be automated (see Appendix A), but we consider both sample sizes as tuning parameters for now. Because the SSD framework with linear proxy functions is substantiated using asymptotic results, there may be bias in the (finite) sample size recommendations. While good empirical performance has been observed in many settings \citep{hagar2025economical, hagar2025precision}, the impact of the tuning parameters on the bias and variance of the resulting sample size recommendations has not been formally studied or compared to alternative SSD approaches. We begin this investigation in Section \ref{sec:mcsd}. 

\subsection{The Robbins-Monro Algorithm}\label{sec:prelim.rm}

The SSD framework that we consider involving sampling distribution exploration is based on the RM algorithm. In the context of SSD, we suppose that we have a function $H(n)$ of the sample size. The relevant function $H(n)$ is the power curve for power-based SSD. For precision-based SSD, we let $H(n)$ be the length probability curve \citep{hagar2025precision}: the probability of satisfying the length criterion in (\ref{eq:prec_prob}) as a function of $n$. The RM algorithm is used to find the root $n^*$ of the equation $H(n) = q$ (or $H(n) = 1 - \beta$ in the case of power). In practice, $n^*$ is not an integer, so the sample size recommendation is $\lceil n^* \rceil$. We discuss the impact of the integer restriction later in this subsection. 

The RM algorithm is useful when it is not possible to observe $H(n)$, but we can obtain instances of a random variable $\tilde{H}(n)$ such that $\mathbb{E}[\tilde{H}(n)] = H(n)$. For SSD, we obtain an instance of $\tilde{H}(n)$ by simulating a single sample $\boldsymbol{y}_n$ from (\ref{eq:prior_pred}); we let $\tilde{H}(n) = 1$ if $\Pr(H_1~|~\boldsymbol{y}_n) \ge \gamma$ or $L_{1-\alpha}(\boldsymbol{y}_n) \le l$, and 0 otherwise. For a continuous input, the RM algorithm generates iterates of the form
\begin{equation}\label{eq:rm}
    n_{r+1} = n_r - a_r \left(\tilde{H}(n_r) - q\right),
\end{equation}
where $r$ indexes the iteration and $a_1, a_2, \dots$ is a sequence of positive step sizes. In that setting, $n_r$ converges in probability to a unique root $n^*$ under the conditions listed in Appendix A.3 that pertain to $\tilde{H}(n)$, $H(n)$, and the sequence $\{a_r\}_{r=0}^{\infty}$ \citep{blum1954approximation}. 

While the RM algorithm can attain a convergence rate of $\mathcal{O}(r^{-1/2})$ \citep{chung1954stochastic}, the optimal $\{a_r\}_{r=0}^{\infty}$ that achieves this convergence depends on $n^*$ and other quantities that are unknown prior to running the algorithm. The convergence of the RM algorithm is also sensitive to the choice of $\{a_r\}_{r=0}^{\infty}$. RM inefficiently converges to $n^*$ if $\{a_r\}_{r=0}^{\infty}$ decreases to 0 too slowly, whereas RM may stop before converging to $n^*$ if $\{a_r\}_{r=0}^{\infty}$ decreases too quickly. To promote more robust convergence without knowledge of unknown quantities, one may use RM with Polyak-Ruppert averaging \citep{polyak1990new, ruppert1988efficient} over longer steps to estimate the root as $\bar{n}_{r} = r^{-1}\sum_{s=1}^r n_s$. \citet{polyak1992acceleration} recommended using a sequence of the form $a_r = c/r^{b}$ defined by constants $c > 0$ and $b \in (0.5, 1)$. The tuning parameters in this framework are thus the initial sample size $n_0$ as well as the constants $c$ and $b$. In practice, we also require a finite bound $m$ on the number of iterations.

The standard results surrounding the convergence of the RM algorithm pertain to the case where $n$ is continuous. For SSD, the iterates in (\ref{eq:rm}) must be rounded to the nearest integer after each iteration to simulate the next sample $\boldsymbol{Y}_n$. While the RM algorithm has been used for SSD, formal investigation of how this rounding impacts convergence or potential bias in the sample size recommendations is limited. The integer restriction also allows for bolstering \citep{chalmers2024solving}. If $n_r = n_s$ for some $s \in \{0, \dots, r-1\}$, the $\tilde{H}(n_s)$ values for such $s$ can be combined with $\tilde{H}(n_r)$ to get a more precise estimate for power or the probability of satisfying the length criterion when computing $n_{r+1}$ via  (\ref{eq:rm}). Pseudocode for the RM algorithm with Polyak-Ruppert averaging and bolstering is provided in Appendix A.3. Since bolstering is not possible when $n$ is continuous, this process may further impact the convergence properties of the RM algorithm. We revisit these considerations when formally comparing this framework based on sampling distribution exploration to that based on sampling distribution estimation.  

\subsection{Benchmark Methods}\label{sec:prelim.det}

Two common SSD approaches based on sampling distribution estimation serve as benchmark methods in this paper. The first benchmark method leverages binary search to deterministically search the $n$-space for the minimum suitable sample size. If $W$ is an upper bound for the sample size, this method requires that we obtain sampling distribution estimates of the relevant posterior summary at $\log_2(W)$ values of $n$. The second benchmark method increases $n$ by regular increments (e.g., by 1, 10, 100, etc.) until power or the probability of satisfying the length criterion based on the estimated sampling distribution is sufficiently large. The required number of sampling distribution estimates depends on the magnitudes of the increment and the minimum suitable sample size. These two SSD methods are less efficient than those based on linear proxy functions since 
more than two sampling distribution estimates are required; hence, neither of these approaches are our main SSD method within the framework based on sampling distribution estimation. Nevertheless, these benchmark methods help contextualize the bias and variance associated with the sample size recommendations obtained using the main SSD methods in Sections \ref{sec:prelim.lin} and \ref{sec:prelim.rm}.

\section{Monte Carlo Sampling Distributions of Sample Size Recommendations}\label{sec:mcsd}

\subsection{Design with Linear Proxy Functions}\label{sec:mcsd.lin}

In this section, we derive the MCSDs of sample size recommendations across repeated implementations of the SSD methods introduced in Section \ref{sec:prelim}. We denote the sample size recommendation for one implementation of a simulation-based SSD method as $n^*$. Across repeated implementations, this sample size recommendation is a random variable $N^*$. The MCSD is defined such that the value of its cumulative distribution function (CDF) at a sample size of $n$ coincides with $\Pr(N^* \le n)$. Through both theoretical consideration and numerical estimation of the MCSD, we can compare the performance of the SSD methods based on sampling distribution estimation and exploration. 

We now prove that the MCSD is approximately normal for the SSD framework based on linear proxy functions, starting with power-based SSD under the conditional approach. In this setting, the order statistics of the sampling distribution estimates at $n_0$ and $n_1$ are used to fit linear proxy functions. Those functions have a limiting slope that is a function of the effect size, which is constant for all $\boldsymbol{y}_n \sim p(\boldsymbol{y}_n)$ under the conditional approach. The sample size recommendation is therefore the value of $n$ such that the line modeling the $\lfloor m \beta \rfloor^{\text{th}}$ order statistic (i.e., the $\beta$-quantile) of the sampling distribution of $\text{logit}(\Pr(H_1~|~\boldsymbol{y}_n))$ is equal to $\text{logit}(\gamma)$. We let $\hat{\xi}$ denote an estimator for the $\beta$-quantile of the sampling distribution of $\text{logit}(\Pr(H_1~|~\boldsymbol{y}_n))$;  we also let $J(\cdot)$ and $j(\cdot)$ respectively denote the CDF and probability density function (PDF) of this distribution. For a large enough number of Monte Carlo iterations $m$, it follows by Bahadur's representation that the approximate distribution of  $\hat{\xi}$ is 
$$\mathcal{N}\left(\xi, \dfrac{\beta(1-\beta)}{j^2(\xi)m}\right),$$ 
where $\xi = J^{-1}(\beta)$ is the true $\beta$-quantile of the sampling distribution. We henceforth add a subscript of 0 or 1 to $\xi$ and $\hat{\xi}$ to distinguish between these quantities when the sample size is $n_0$ or $n_1$.

    Based on the linear proxy functions, the sample size recommendation $N^*$ is the value of $n$ for which the line passing through $(n_0, \hat{\xi}_0)$ and $(n_1, \hat{\xi}_1)$ has a $y$-value of $\text{logit}(\gamma)$. This sample size is
    $$N^* = k(\hat{\xi}_0,\hat{\xi}_1) = \dfrac{(\text{logit}(\gamma) - \hat{\xi}_0)(n_1 - n_0)}{\hat{\xi}_1 - \hat{\xi}_0} + n_0.$$
    Thus, $N^*$ is a continuous function $k(\cdot)$ of the independent normal random variables $\hat{\xi}_0$ and $\hat{\xi}_1$. By the delta method, the MCSD for $N^*$ is approximately 
    \begin{equation}\label{eq:mcsd.pwr}
    \mathcal{N}\left(k(\xi_0, \xi_1), \dfrac{\partial k}{\partial \boldsymbol{\xi}}^{\text{T}}\boldsymbol{\Sigma}\dfrac{\partial k}{\partial \boldsymbol{\xi}}\right),
    \end{equation}
    where $\boldsymbol{\xi} = (\xi_0, \xi_1)$ and $\boldsymbol{\Sigma}$ is the covariance matrix of $(\hat{\xi}_0, \hat{\xi}_1)$ that follows from Bahadur's representation. We note that $k(\xi_0, \xi_1)$ depends on $n_0$ and $n_1$, which are not random when these sample sizes are treated as tuning parameters.

    While the MCSD in (\ref{eq:mcsd.pwr}) is not computed in practice, we can draw valuable insights from this result. First, the framework from \citet{hagar2025economical} prompts \emph{asymptotically} consistent sample size recommendations. However, unsuitable choices for $n_0$ and $n_1$ could give rise to material bias in $N^*$ since the sample size recommendation is necessarily finite. As discussed in Appendix A of the supplement and Section \ref{sec:num}, the second sample size $n_1$ can be automatically chosen to mitigate the potential for bias in the MCSD. Because the MCSD is approximately normal, it is reasonable to use standard metrics like the bias, variance, and mean squared error (MSE) of the MCSD with respect to the true minimum sample size, $H^{-1}(1 - \beta)$, to quantify the performance of an SSD method. We extend the result for the MCSD in (\ref{eq:mcsd.pwr}) to accommodate the predictive approach and empirically verify the MCSD approximations using simulation in Appendix B.1. 

    In Appendix B.2 of the supplement, we derive MCSD approximations similar to that in (\ref{eq:mcsd.pwr}) for precision-based SSD with linear proxy functions. Under the conditional approach, the MCSD for $N^*$ is approximately 
    \begin{equation}\label{eq:mcsd.prec}
    \mathcal{N}\left(k(\boldsymbol{\xi}), \dfrac{\partial k}{\partial \boldsymbol{\xi}}^{\text{T}}\boldsymbol{\Sigma}\dfrac{\partial k}{\partial \boldsymbol{\xi}}\right),
    \end{equation}
    where $\xi_{0.5, n}$ and $\xi_{q, n}$ now represent the median and $q$-quantile of the sampling distribution of $\log(L_{1-\alpha}(\boldsymbol{y}_n))$, $\boldsymbol{\xi} = (\xi_{0.5, n_0}, \xi_{q, n_0}, \xi_{0.5, n_1}, \xi_{q, n_1})$, and $\boldsymbol{\Sigma}$ is the covariance matrix of $(\hat{\xi}_{0.5, n_0}, \hat{\xi}_{q, n_0}, \hat{\xi}_{0.5, n_1}, \hat{\xi}_{q, n_1})$ that follows from Bahadur's representation. We provide the form of the function $k(\cdot)$ for precision-based SSD and extend this result to the predictive approach in Appendix B.2. The mean of the MCSD in (\ref{eq:mcsd.prec}) again depends on the values of $n_0$ and $n_1$, so the considerations for power-based SSD regarding potential bias in $N^*$ and automating the choice of $n_1$ to mitigate this bias remain relevant. 

\subsection{The Robbins-Monro Algorithm}\label{sec:mcsd.rm}

For various implementations of the RM algorithm, the distribution of the algorithm output $N^*$ as the number of Monte Carlo iterations $m \rightarrow \infty$ has been derived (see e.g., \citet{fabian1968asymptotic} for consideration of the standard RM algorithm). While these distributions of the algorithm output can be framed as MCSDs, this nomenclature is not used in the stochastic root-finding literature. We also note that existing derivations of these asymptotic distributions have pertained to the case where the function $H(n)$ is continuous.

We focus on the approximate distribution of $N^*$ when the RM algorithm is implemented with Polyak-Ruppert averaging. By directly applying results from \citet{polyak1992acceleration}, the resulting MCSD of $N^*$  is
\begin{equation}\label{eq:mcsd.rm}
    \mathcal{N}\left(n^*, \dfrac{H(n^*)[1-H(n^*)]}{[H^\prime(n^*)]^2m} \right),
    \end{equation}
where $H(n)$ is a continuous analog to the power or length probability curve, and $n^*$ is the sample size at which $H(n)$ equals $1 - \beta$ or $q$. The $H(n^*)[1-H(n^*)]$ component of the variance in (\ref{eq:mcsd.rm}) comes from the variance of the observable binary $\tilde{H}(n)$ instances near the root. For a continuous input $n$, $N^*$ should thus converge to $n^*$ as $m \rightarrow \infty$, and the variance of the MCSD in (\ref{eq:mcsd.rm}) should not depend on the values of $c > 0$ and $b \in (0.5, 1)$ that define the sequence of step sizes. However, formal study of the MCSD of the algorithm output $N^*$ is limited when the input is restricted to be an integer as in SSD. 

To address this gap, our numerical studies in Section \ref{sec:num} consider both standard, integer-restricted SSD and a continuous analog to SSD, which allows us to consider the impact of rounding on the MCSD in isolation. These simulations will assess how the values of the tuning parameters -- $n_0$, $b$, and $c$ -- impact the MCSD in practice, particularly for the version of the RM algorithm with rounding. Again, the approximate normality of the MCSD in (\ref{eq:mcsd.rm}) for continuous $H(n)$ suggests that it is logical to use the bias, variance, and MSE of the MCSD to quantify the performance of the SSD methods based on sampling distribution exploration.

\subsection{Benchmark Methods}\label{sec:mcsd.det}

Lastly, we define the MCSD of $N^*$ for the two benchmark methods introduced in Section \ref{sec:prelim.det}. We now introduce general notation for power-based SSD that applies when either binary search or incrementally increasing the sample size is used to deterministically search the $n$-space. We let $\mathcal{P}$ denote a set of paths to search the $n$-space such that all paths start at the same initial sample size $n_0$. The subsequent sample sizes on the path $s = \{n_0, n_{s,1}, \dots, n_{s, t_s} \} \in \mathcal{P}$ depend on estimated sampling distributions of posterior summaries, each obtained using $m$ Monte Carlo iterations. The length $t_s$ of the path may differ across $s \in \mathcal{P}$. 

We let $\bar{H}(n)$ denote the estimate of power based on a sampling distribution estimate at $n$. For a general path $s$, the next sample size $n_{s, 1}>n_0$ if $\bar{H}(n_0) < 1 - \beta$, and $n_{s, 1} < n_0$ otherwise. The probability that $\bar{H}(n) > 1 - \beta$ is approximately 
\begin{equation}\label{eq:det}
1 - \Phi\left( \sqrt{m} \times \dfrac{(1 - \beta) - H(n)}{\sqrt{H(n)(1 - H(n))}}\right).
\end{equation}
As $m \rightarrow \infty$, the probability in (\ref{eq:det}) approaches 1 if $H(n) > 1 - \beta$, and 0 if $H(n) < 1 - \beta$. It is, however, often computationally prohibitive to use very large $m$ when the sampling distribution of posterior summaries must be estimated at many values of $n$. In practice, the probability in (\ref{eq:det}) may not approximate 0 or 1 with moderate $m$ when $H(n) \approx 1 - \beta$, and there are thus many potential paths in $\mathcal{P}$ that we may take to search the $n$-space. We denote the path taken as a random variable $\mathcal{S}$, where the probability $\Pr(\mathcal{S} = s)$ depends on the initial sample size $n_0$, the value of $m$ per sampling distribution estimate, and the method used to search the $n$-space. 

To define the MCSD, we introduce a set of paths $\mathcal{P}_n$ for which the sample size of $n$ is recommended. It follows that $$\mathcal{P}_n = \left\{ s \in \mathcal{P}: n \in s ~\cap~ \bar{H}(n) \ge 1 - \beta ~ \cap ~\{ \bar{H}(n^-) < 1 - \beta ~\forall ~n^- \in s: n^- < n\}\right\}.$$
In particular, the sample size $n$ must be considered on the path $s \in \mathcal{P}$, $\bar{H}(n)$ must be at least $1 - \beta$, and $\bar{H}(n^-)$ at any smaller sample size $n^-$ on the path must be less than $1 - \beta$. For precision-based SSD, the logic and notation above applies when the target probability $q$ is used instead of the target power. The MCSD of $N^*$ is formally defined such that 
\begin{equation}\label{eq:mcsd.det}
\Pr(N^* = n) = \sum_{s \in \mathcal{P}_n}\Pr(\mathcal{S} = s).\end{equation}
There is no guarantee that the MCSD in (\ref{eq:mcsd.det}) is approximately normal as $m \rightarrow \infty$. In the numerical studies in Section \ref{sec:num}, it is therefore useful to visualize the estimated MCSDs to assess how the initial sample size $n_0$ and method used to search the $n$-space impact the performance of the benchmark SSD methods.

\section{Numerical Studies}\label{sec:num}

\subsection{Design with Power Criteria}\label{sec:num.pwr}

The numerical studies in this paper consider an example that is motivated by a clinical trial for semaglutide development \citep{wilding2021once}. In this clinical trial, patients were respectively given a weekly semaglutide injection or placebo for 68 weeks. One primary outcome in that trial was the percentage change in body weight over the course of the study. We suppose that we want to design a similar comparison based on the following linear regression model: $y_i = \psi_0 + \psi_1x_{1i} + \psi_2x_{2i} + \varepsilon_i$, where $y$ is percentage change in body weight, $x_1 \in \{0,1\}$ denotes if the patient received the semaglutide, $x_2$ is the patient's baseline waist circumference in centimeters, and $\varepsilon_i \sim \mathcal{N}(0, \sigma^2_{\varepsilon})$ are independent error terms. For this example, Bayesian analysis is facilitated using a conjugate normal-inverse-gamma prior for $\boldsymbol{\eta} = (\boldsymbol{\psi}, \sigma^2_{\varepsilon})$. 

While this regression model is relatively simple, it offers two advantages for our numerical studies. First, the use of conjugate priors makes it computationally feasible to conduct hundreds of thousands of sample size calculations. Second, this model allows us to consider an analog to SSD that is continuous in $n$. We let $n$ be the sample size per treatment for this example such that the total sample size is $2n$. The log-likelihood function for the model is thus
\begin{align}\label{eq:lik}
      l(\boldsymbol{\eta}; \boldsymbol{y}, {\bf{X}}) &= -n\log(2\pi\sigma^2) - \dfrac{1}{2\sigma^2}\sum_{i=1}^{2n}(y_i - \psi_0 - \psi_1x_{1i} -\psi_2x_{2i})^2. 
\end{align}
To compute the log-likelihood in (\ref{eq:lik}), we do not require individual-level data $(y_i, x_{1i}, x_{2i})$ for patients $i = 1, \dots, 2n$. Along with the treatment-specific $n$, the sums $\sum_{i=1}^{2n} y_i$, $\sum_{i=1}^{2n} y_i^2$, $\sum_{i=1}^{2n} y_ix_{1i}$, $\sum_{i=1}^{2n} y_ix_{2i}$, $\sum_{i=1}^{2n} x_{2i}$, $\sum_{i=1}^{2n} x_{2i}^2$,   and $\sum_{i=1}^{2n} x_{2i}x_{1i}$ are sufficient statistics for the likelihood and corresponding posterior of $\boldsymbol{\eta}$. Under the linear model described above, we show in Appendix C.1 of the supplement that all required sufficient statistics can be simulated using normal and chi-square random variables -- even when $n$ is non-integer. This continuous SSD approach based on simulating sufficient statistics is not available for more complex examples, but we use it to assess the performance of the RM algorithm. All SSD methods except the continuous RM variant can trivially accommodate design with more statistically complex models.

We first consider power-based SSD under the conditional approach. The degenerate design prior $p_D(\boldsymbol{\eta})$ is such that $\boldsymbol{\psi} = (-25.75, 7, 0.25)$ and $\sigma_{\varepsilon} = 10.07$ are used to generate $\boldsymbol{y}_n$ in all Monte Carlo iterations. We also assume that $\{x_{2i}\}_{i = 1}^{2n} \overset{\text{i.i.d.}}{\sim} \mathcal{N}(115, 14.5^2)$. The target of inference is $\theta = \psi_1$, the increased amount of weight loss (in \%) associated with taking the semaglutide injections. We aim to support the hypothesis $H_1: \theta \in (\delta_L, \delta_U) = (5, \infty)$, which would suggest the semaglutide yields substantial weight loss of at least 5\% more than the placebo to offset treatment side effects. We use a decision threshold of $\gamma = 0.95$ to approximately attain a type I error rate of 5\%. The target power for this example is $1 - \beta = 0.8$. In this subsection, we use a diffuse normal-inverse-gamma prior  such that $\sigma^2_{\varepsilon} \sim \text{InverseGamma}(1, 1)$ and $\boldsymbol{\psi}~|~\sigma^2_{\varepsilon} \sim \mathcal{N}({\bf{0}}_{3}, 100 \times \mathbb{I}_{3})$, where $\mathbb{I}_{3}$ is the $3 \times 3$ identity matrix. Based on computationally intensive simulations, the true minimum sample size is $n = 314$.

\begin{figure}[!tb] \centering 
		\includegraphics[width = \textwidth]{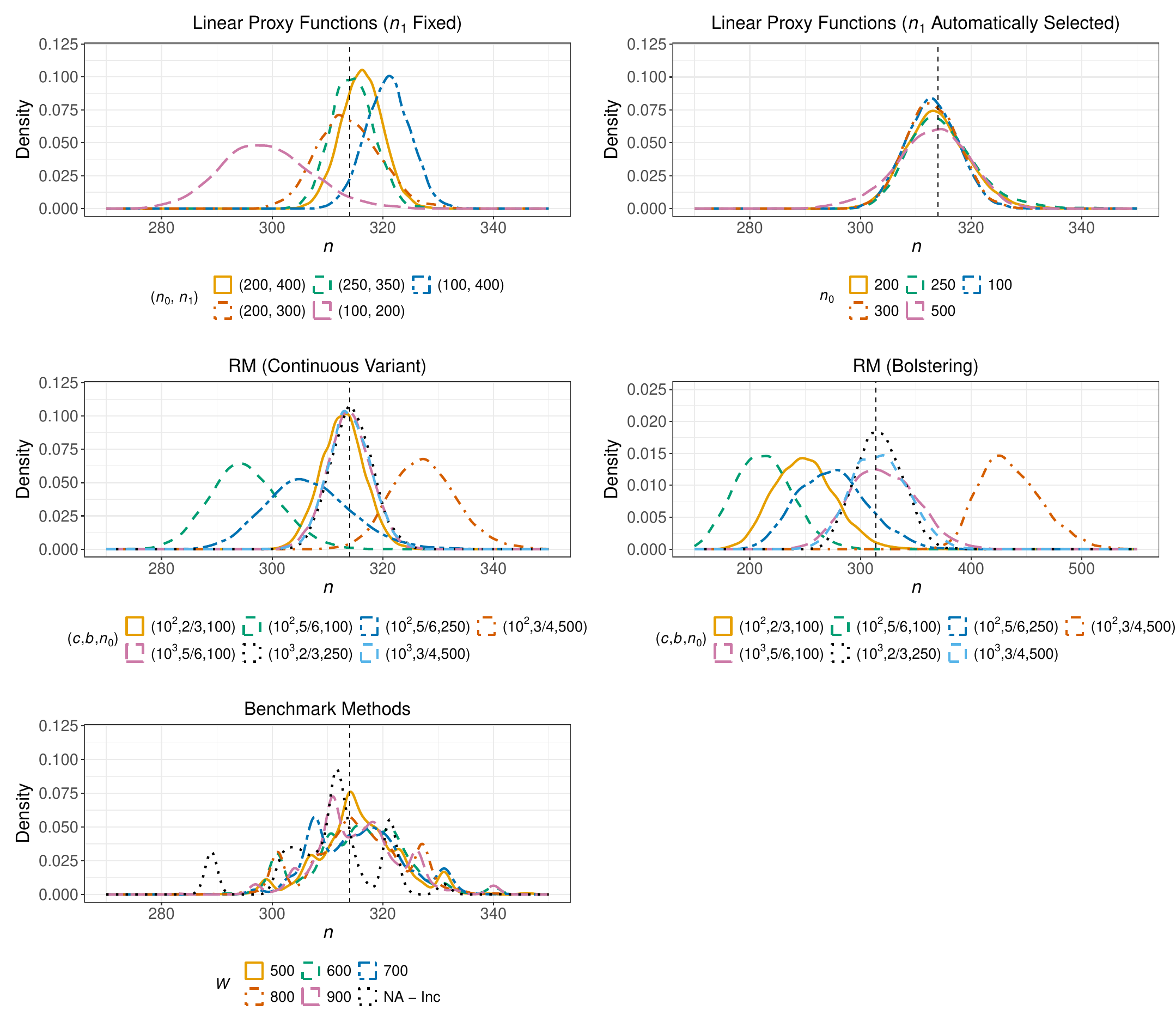} 

		\caption{\label{fig:pwr.cond} Empirical MCSDs obtained using various power-based SSD methods under the conditional approach. The tuning parameters for each method are listed in the legend. The true sample size recommendation of $n = 314$ is given by the vertical dashed line.} 
	\end{figure}

   We now consider the SSD approach based on linear proxy functions under various settings for the tuning parameters ($n_0$ and $n_1$). For each setting, we implemented the sample size calculation $5 \times 10^3$ times to empirically estimate the MCSD, where each of the two sampling distribution estimates was constructed using $m = 5\times10^3$ iterations. The results  are visualized in the top left plot of Figure \ref{fig:pwr.cond}. We observe that there can be substantial bias in the MCSD if neither $n_0$ or $n_1$ are near the true sample size recommendation. This result is not surprising. The linear proxy functions are (globally) suitable based on asymptotic theory, but they may only be locally suitable for a range of finite sample sizes near $n_0$ or $n_1$ in practice. To mitigate the bias in the MCSD, the second sample size $n_1$ can be automatically selected using the limiting slopes from \citet{hagar2025economical}. The top right plot of Figure \ref{fig:pwr.cond} shows that there is minimal bias in the resulting MCSD under a broad range of values for $n_0$. 

      We then investigated the two variants of the RM algorithm with Polyak-Ruppert averaging. For both variants, we estimate the MCSD for each setting by implementing $5 \times 10^3$ sample size calculations, each with $m = 10^4$ iterations.  The same number of total iterations are thus used for all sample size calculations based on the RM algorithm and the linear proxy functions. For both RM variants, the first $10^3$ iterations of $\{n_r\}_{r=1}^{10^4}$ were excluded as burn-in before taking the average of the final $9 \times 10^3$ sample sizes values as the recommended $n$. The first RM variant is such that $n$ is continuous. We considered extensive combinations of the tuning parameters ($n_0$, $c$, and $b$), and the results for seven of these settings are visualized in the middle left plot of Figure \ref{fig:pwr.cond}. For the settings where $c = 100$, the step sizes are not always large enough to ensure convergence to $n = 314$ with only $m = 10^4$ iterations; trace plots of $\{n_r\}_{r=1}^{m}$ in Appendix C.2 of the supplement confirm that these sequences are still approaching $n = 314$. For the settings with larger $c$ values, there is minimal bias in the MCSD and little sensitivity to the values of the other tuning parameters, as suggested in (\ref{eq:mcsd.rm}).

    The second RM variant with bolstering is such that the value for $n$ was rounded to the nearest integer in each iteration. We visualize the MCSDs for the same seven $(n_0, c, b)$ combinations in the middle right plot of Figure \ref{fig:pwr.cond}. This subplot has different scales for its axes than the others. For the settings where $c = 100$, this version of RM generally does not converge to the correct sample size. For the settings with larger $c$ values, we see better convergence with slight positive bias in the MCSD due to rounding. The variance of these MCSDs is much greater than those based on the continuous RM variant. Unlike with the continuous RM variant, increasing $m$ may not lead to better convergence. The magnitude of the step sizes eventually decreases below 0.5, and the next value of $n$ rounds back to its previous value in subsequent iterations as illustrated in the trace plots in Appendix C.2. Therefore, the RM algorithm with bolstering may appear to converge at a value that is not near the true minimum $n$. 

We also consider the benchmark methods for reference. For binary search, we considered five values for the tuning parameter of the upper bound: $W \in \{500, 600, \dots, 900\}$. We also considered an approach in which the sampling distribution was estimated while incrementally increasing the value of $n$ (by hundreds then by tens and ones) to find the smallest $n$ where the power estimate was at least $1 - \beta$. There is no tuning parameter for this method. Because $m = 10^3$ iterations were used per sampling distribution estimate, the runtime of binary search with $W = 900$ is similar to that of SSD with linear proxy functions. The SSD method based on incremental increases is slower than binary search. The bottom left plot of Figure \ref{fig:pwr.cond} visualizes the results based on $5\times10^3$ sample size calculations per MCSD. The resulting MCSDs are not approximately normal; the sample size recommendations generally clump together based on whether the power estimates in early iterations exceed $1 - \beta$. The MCSD variance under binary search is greater than under the approach with linear proxy functions. There is also negative bias for SSD based on incremental increases of $n$: we are likely to stop early after obtaining a higher-than-expected power estimate at one of the successive sample sizes considered.

Table \ref{tab:pwr.cond} numerically summarizes the results for the best and worst performing settings for each SSD method considered in Figure \ref{fig:pwr.cond}. Best and worst performance is based on the root mean squared error (RMSE) of the MCSD with respect to the true sample size recommendation of $n = 314$; the bias and standard deviation (SD) of the MCSD are also reported. Table \ref{tab:pwr.cond} confirms that there is a bias-variance trade-off under the two SSD methods with linear proxy functions. When $n_0$ and $n_1$ are both fixed at sample sizes relatively close to the true minimum $n$, the MCSD variance is smaller, but unsuitable choices for both $n_0$ and $n_1$ can lead to bias in the MCSD. The RMSE of the MCSD is generally smallest for the continuous RM variant when the sequence converges to the correct sample size; however, this method cannot be used for settings where sufficient statistics are unavailable, and it is not trivial to verify correct convergence. The SD of the RM variant with bolstering for this example could be made smaller by choosing larger step sizes, but it is not straightforward to choose the constant $c$ prior to implementation. We also note that the RMSE for the best performing benchmark method is greater than the RMSE of the worst performing setting with linear proxy functions when $n_1$ is automatically selected. In Appendix C.3 of the supplement, we summarize comparable results based on similar simulations for power-based SSD under the predictive approach.

\begin{table}[!htb]
\centering
\caption{Summary of the RMSE, bias, and SD of the MCSD in the best and worst performing settings for each SSD method from Figure \ref{fig:pwr.cond}.}
\label{tab:pwr.cond}
\begin{tabular}{lcccccccc}
                           & \multicolumn{4}{c}{Best Performing Setting}              & \multicolumn{4}{c}{Worst Performing Setting}               \\ \cline{2-9} 
    \multicolumn{1}{l|}{}                       & Parameters                           & RMSE  & Bias  &  \multicolumn{1}{c|}{SD}    & Parameters                           & RMSE   & Bias   & SD    \\ \cline{2-9} 
\multicolumn{1}{l|}{Linear (Fixed)}    & $(250, 350)$         & 3.88  & 0.55  & \multicolumn{1}{c|}{3.84}  & $ (100, 200)$         & 17.16  & -15.12 & 8.12  \\
\multicolumn{1}{l|}{Linear  (Selected)} & $100$                       & 4.87  & -1.07 & \multicolumn{1}{c|}{4.75}  & $500$                       & 6.72   & -1.11  & 6.63  \\
\multicolumn{1}{l|}{RM (Continuous)}            & $ (10^3, 2/3, 250) $ & 3.76  & 0.37  & \multicolumn{1}{c|}{3.75}  & $(10^2, 5/6, 100) $ & 20.00     & -18.99 & 6.27  \\
\multicolumn{1}{l|}{RM (Bolstering)}            & $(10^3, 2/3, 250) $ & 20.84 & 2.66  & \multicolumn{1}{c|}{20.68} & $(10^2, 3/4, 500) $ & 125.05 & 121.90  & 27.91 \\
\multicolumn{1}{l|}{Benchmark}                  & $ 500$                         & 7.63  & 1.69  & \multicolumn{1}{c|}{7.44}  & Inc                               & 10.04  & -3.85  & 9.27 
\end{tabular}
\end{table} 

\subsection{Design with Precision Criteria}\label{sec:num.prec}

We next consider precision-based SSD under the predictive approach. In this subsection, we modify the linear regression example from Section \ref{sec:num.pwr}, corresponding to the log-likelihood in (\ref{eq:lik}). Our objective of this experiment is to obtain a $100 \times (1- \alpha) = 95\%$ CI for $\theta = \psi_1$ that has length at most $l = 4$ with target probability $q = 0.9$. For illustration, we now use a normal-inverse-gamma prior that is relatively informative for $\theta$  such that $\sigma^2_{\varepsilon} \sim \text{InverseGamma}(7, 50)$ and $\boldsymbol{\psi}~|~\sigma^2_{\varepsilon} \sim \mathcal{N}((0, 6, 0), (100, 9, 100) \times \mathbb{I}_{3})$. 

\begin{figure}[!tb] \centering 
		\includegraphics[width = \textwidth]{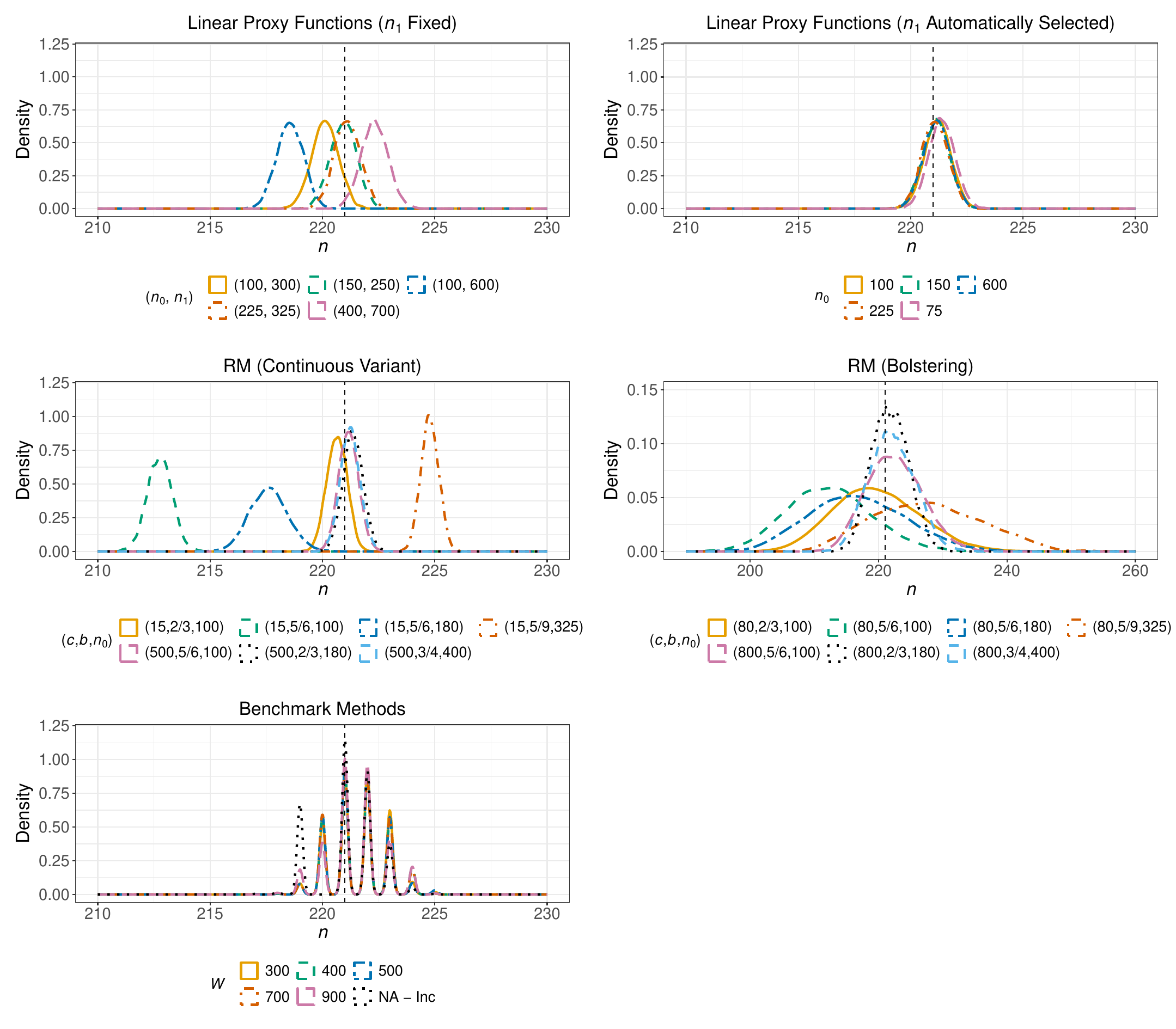} 

		\caption{\label{fig:prec.pred} Empirical MCSDs obtained using various precision-based SSD methods under the predictive approach. The tuning parameters for each method are listed in the legend. The true sample size recommendation of $n = 221$ is given by the vertical dashed line.} 
	\end{figure}

Under the predictive approach, the value of $\sigma_{\varepsilon}$ used for data generation is no longer held fixed at a value of 10.07 for all Monte Carlo iterations. Instead, we define a design prior $p_D(\boldsymbol{\eta})$ such that $\sigma_{\varepsilon}$ is drawn according to a $ \mathcal{U}[9, 11]$ distribution in each iteration. We do not accommodate uncertainty in the coefficients $\boldsymbol{\psi}$ at the data-generating stage since the inverse Fisher information of $\theta$ is only a function of $\sigma_{\varepsilon}$ along with the covariates $x_1$ and $x_2$. The true minimum sample size is $n = 221$ based on computationally intensive simulations. We considered the same five SSD methods as in Section \ref{sec:num.pwr}, but we modified several of the tuning parameters as recorded in Figure \ref{fig:prec.pred}. Once again, the axes for the middle right plot for the RM variant with bolstering differ from the axes for the other subplots. For each SSD method, we estimated the MCSD by implementing $5\times10^3$ sample size calculations under each setting for the tuning parameters; the number of Monte Carlo iterations used for each method is the same as in Section \ref{sec:num.pwr}. 

    Table \ref{tab:prec.pred} numerically summarizes the results for the best and worst performing settings for each SSD method considered in Figure \ref{fig:prec.pred}. We draw similar conclusions about the performance of the various SSD methods based on Figure \ref{fig:prec.pred} and Table \ref{tab:prec.pred} as for power-based SSD under the conditional approach in Section \ref{sec:num.pwr}. Nevertheless, we briefly elaborate on the differences between the two numerical studies. First, the MCSD variances are generally smaller for precision-based SSD. This variance reduction occurs because $H^{\prime}(n^*)$ is usually larger in this setting: whereas the power criterion requires that the posterior distribution of $\theta$ concentrates in the interval $(\delta_L, \delta_U)$, the length criterion only requires that this distribution concentrates in any interval of length $l$. The impact of $H^{\prime}(n^*)$ on the MCSD variance for the continuous RM variant is explicitly clear in (\ref{eq:mcsd.rm}).  

    \begin{table}[!htb]
\centering
\caption{Summary of the RMSE, bias, and SD of the MCSD in the best and worst performing settings for each SSD method from Figure \ref{fig:prec.pred}.}
\label{tab:prec.pred}
\begin{tabular}{lcccccccc}
                           & \multicolumn{4}{c}{Best Performing Setting}              & \multicolumn{4}{c}{Worst Performing Setting}               \\ \cline{2-9} 
    \multicolumn{1}{l|}{}                       & Parameters                           & RMSE  & Bias  &  \multicolumn{1}{c|}{SD}    & Parameters                           & RMSE   & Bias   & SD    \\ \cline{2-9} 
\multicolumn{1}{l|}{Linear (Fixed)}    & $(225, 325)$         & 0.59  & 0.11  & \multicolumn{1}{c|}{0.58}  & $ (100, 600)$         & 2.52  & -2.45 & 0.59  \\
\multicolumn{1}{l|}{Linear  (Selected)} & $225$                       & 0.59  & 0.08 & \multicolumn{1}{c|}{0.58}  & $75$                       & 0.69   & 0.38  & 0.58  \\
\multicolumn{1}{l|}{RM (Continuous)}            & $ (500, 5/6, 100) $ & 0.47  & 0.17  & \multicolumn{1}{c|}{0.44}  & $(15, 5/6, 100) $ & 8.26     & -8.24 & 0.55  \\
\multicolumn{1}{l|}{RM (Bolstering)}            & $(800, 2/3, 180) $ & 2.92 & 0.78  & \multicolumn{1}{c|}{2.81} & $(80, 5/6, 100) $ & 10.88 & -8.69  & 6.54 \\
\multicolumn{1}{l|}{Benchmark}                  & $ 500$                         & 1.31  & 0.53  & \multicolumn{1}{c|}{1.20}  & $400$                               & 1.37  & 0.57  & 1.24 
\end{tabular}
\end{table} 

This larger value of $H^{\prime}(n^*)$ for this example leads to better performance for the benchmark methods. However, the benchmark methods are still outperformed by SSD with linear proxy functions when $n_1$ is automatically selected, which offers the best robust performance. Because $q = 0.9$ is much closer to 1 than 0, the continuous variant of the RM algorithm converges asymmetrically with respect to the initial sample size $n_0$. For sample sizes $n_r >> n^*$ such that $\Pr(\tilde{H}(n_r) = 1) \approx 1$, the step size in (\ref{eq:rm}) is $0.1 \times a_r$. In contrast, the step size is in (\ref{eq:rm}) is $0.9\times a_r$ when $n_r << n^*$ such that $\Pr(\tilde{H}(n_r) = 0) \approx 1$. Good convergence to $n^*$ is still possible when the tuning parameters for the continuous RM variant are suitable. In Appendix C.4 of the supplement, we summarize comparable results based on similar simulations for precision-based SSD under the conditional approach.

\section{Discussion}\label{sec:disc}

 In this paper, we assessed two distinct strategies for Bayesian SSD based on Monte Carlo simulations. One strategy involved estimating entire sampling distributions of posterior summaries at various sample sizes $n$ to find the minimum suitable sample size. The other approach explored these sampling distributions using stochastic root finding over the $n$-space, where the entire sampling distribution of posterior summaries is not estimated at any given sample size. Both strategies can be used for SSD based on power or precision criteria, with the option to incorporate uncertainty in the data-generating parameters by way of a design prior. 
 
 We compared exemplar methods for the two strategies, based on linear proxy functions and the RM algorithm, to benchmark SSD methods. To quantify the performance of each SSD method, we defined the Monte Carlo sampling distribution of sample size recommendations; we also proved that the MCSD is approximately normal for the SSD approach with linear proxy functions under various conditions. Based on the bias, variance, and MSE of the resulting MCSDs, SSD with linear proxy functions in the sampling distribution estimation framework offered the most robust, suitable performance when the second sample size $n_1$ was automatically selected. However, the continuous RM variant in the sampling distribution exploration framework often had the smallest RMSE when the tuning parameters were well selected. This continuous variant cannot be applied with complex statistical models, and there were material convergence issues for the integer-restricted SSD method based on the RM algorithm.     

The SSD approach based on linear proxy functions offers reliable performance because the Monte Carlo simulations are augmented with asymptotic theory. Because the continuous RM variant offered good performance, it may be possible to construct an integer-restricted SSD method in the sampling distribution exploration framework that outperforms all SSD methods considered in this paper. In particular, we could use an RM variant with a (potentially non-decreasing) sequence of step sizes $\{a_r\}_{r=0}^{\infty}$ that is created using the theory associated with the linear proxy functions. While the sample size calculations in the sampling distribution estimation framework may be efficiently parallelized, the sampling distribution exploration framework may be advantageous in that sample size calculations can be terminated early if the sequence $\{n_r\}_{r=0}^m$ remains relatively constant across subsequent iterations. Future research could investigate the performance of new SSD methods and formally assess the benefits of parallelization and early termination.

Finally, we acknowledge that both main SSD methods considered are not ideal for small-sample settings. The use of linear proxy functions is substantiated using large-sample theory. Moreover, a lower bound for the sample size must be selected in practice to implement the RM algorithm, which could introduce bias into the Polyak-Ruppert averaging. We also note that more complex methods for Bayesian SSD exist, especially when SSD is based on power criteria. For instance, the design of Bayesian clinical trials may involve selecting suitable sample sizes and decision thresholds to attain criteria for various operating characteristics in complex adaptive designs. Such sample size calculations can be implemented in the sampling distribution estimation framework (see e.g., \citet{hagar2026group, hagar2026design, hagar2026economical}). However, those design problems are not well suited for sampling distribution exploration based on univariate stochastic root finding since a more complex and unstable multivariate search across sample sizes and decision thresholds would be required.

 \section*{Supplementary Material}
 These materials include formal algorithms for the SSD methods in Section \ref{sec:prelim}, extensions of the results in Section \ref{sec:mcsd.lin} to the predictive approach, and additional simulations. The code to conduct the numerical studies in the paper is available online:  \url{https://github.com/lmhagar/MCSD}.


	\section*{Funding}
 
LH acknowledges the support of a postdoctoral fellowship from the Natural Sciences and Engineering Research Council of Canada (NSERC). PG acknowledges support from NSERC Discovery Grant RGPIN-2025-04355. 
	

\bibliographystyle{chicago}


\begin{thebibliography}{}

\bibitem[\protect\citeauthoryear{Amaratunga}{Amaratunga}{1999}]{amaratunga1999searching}
Amaratunga, D. (1999).
\newblock Searching for the right sample size.
\newblock {\em The American Statistician\/}~{\em 53\/}(1), 52--55.

\bibitem[\protect\citeauthoryear{Berry, Carlin, Lee, and Muller}{Berry et~al.}{2011}]{berry2010bayesian}
Berry, S.~M., B.~P. Carlin, J.~J. Lee, and P.~Muller (2011).
\newblock {\em Bayesian Adaptive Methods for Clinical Trials}.
\newblock CRC press.

\bibitem[\protect\citeauthoryear{Blum}{Blum}{1954}]{blum1954approximation}
Blum, J.~R. (1954).
\newblock Approximation methods which converge with probability one.
\newblock {\em The Annals of Mathematical Statistics\/}~{\em 25\/}(2), 382--386.

\bibitem[\protect\citeauthoryear{Brutti, De~Santis, and Gubbiotti}{Brutti et~al.}{2008}]{brutti2008robust}
Brutti, P., F.~De~Santis, and S.~Gubbiotti (2008).
\newblock Robust bayesian sample size determination in clinical trials.
\newblock {\em Statistics in Medicine\/}~{\em 27\/}(13), 2290--2306.

\bibitem[\protect\citeauthoryear{Brutti, De~Santis, and Gubbiotti}{Brutti et~al.}{2014}]{brutti2014bayesian}
Brutti, P., F.~De~Santis, and S.~Gubbiotti (2014).
\newblock Bayesian-frequentist sample size determination: a game of two priors.
\newblock {\em Metron\/}~{\em 72\/}(2), 133--151.

\bibitem[\protect\citeauthoryear{Chalmers}{Chalmers}{2024}]{chalmers2024solving}
Chalmers, R.~P. (2024).
\newblock Solving variables with monte carlo simulation experiments: A stochastic root-solving approach.
\newblock {\em Psychological Methods\/}.

\bibitem[\protect\citeauthoryear{Chung}{Chung}{1954}]{chung1954stochastic}
Chung, K.~L. (1954).
\newblock On a stochastic approximation method.
\newblock {\em The Annals of Mathematical Statistics\/}~{\em 23}, 463--483.

\bibitem[\protect\citeauthoryear{De~Santis}{De~Santis}{2007}]{de2007using}
De~Santis, F. (2007).
\newblock Using historical data for {B}ayesian sample size determination.
\newblock {\em Journal of the Royal Statistical Society: Series A (Statistics in Society)\/}~{\em 170\/}(1), 95--113.

\bibitem[\protect\citeauthoryear{De~Santis and Pacifico}{De~Santis and Pacifico}{2004}]{santis2004two}
De~Santis, F. and M.~P. Pacifico (2004).
\newblock Two experimental settings in clinical trials: predictive criteria for choosing the sample size in interval estimation.
\newblock In {\em Applied Bayesian Statistical Studies in Biology and Medicine}, pp.\  109--130. Springer.

\bibitem[\protect\citeauthoryear{Fabian}{Fabian}{1968}]{fabian1968asymptotic}
Fabian, V. (1968).
\newblock On asymptotic normality in stochastic approximation.
\newblock {\em The Annals of Mathematical Statistics\/}, 1327--1332.

\bibitem[\protect\citeauthoryear{Golchi and Willard}{Golchi and Willard}{2024}]{golchi2024estimating}
Golchi, S. and J.~J. Willard (2024).
\newblock Estimating the sampling distribution of posterior decision summaries in bayesian clinical trials.
\newblock {\em Biometrical Journal\/}~{\em 66\/}(8), e70002.

\bibitem[\protect\citeauthoryear{Green and MacLeod}{Green and MacLeod}{2016}]{green2016simr}
Green, P. and C.~J. MacLeod (2016).
\newblock {SIMR}: An {R} package for power analysis of generalized linear mixed models by simulation.
\newblock {\em Methods in Ecology and Evolution\/}~{\em 7\/}(4), 493--498.

\bibitem[\protect\citeauthoryear{Grill, Valko, and Munos}{Grill et~al.}{2015}]{grill2015black}
Grill, J.-B., M.~Valko, and R.~Munos (2015).
\newblock Black-box optimization of noisy functions with unknown smoothness.
\newblock {\em Advances in Neural Information Processing Systems\/}~{\em 28}.

\bibitem[\protect\citeauthoryear{Gubbiotti and De~Santis}{Gubbiotti and De~Santis}{2011}]{gubbiotti2011bayesian}
Gubbiotti, S. and F.~De~Santis (2011).
\newblock A bayesian method for the choice of the sample size in equivalence trials.
\newblock {\em Australian \& New Zealand Journal of Statistics\/}~{\em 53\/}(4), 443--460.

\bibitem[\protect\citeauthoryear{Hagar and Golchi}{Hagar and Golchi}{2026}]{hagar2026design}
Hagar, L. and S.~Golchi (2026).
\newblock Design of {B}ayesian clinical trials with clustered data.
\newblock {\em Statistics in Medicine\/}~{\em 45\/}(6-7), e70488.

\bibitem[\protect\citeauthoryear{Hagar, Golchi, and Klein}{Hagar et~al.}{2026}]{hagar2026group}
Hagar, L., S.~Golchi, and M.~B. Klein (2026).
\newblock Group sequential design with posterior and posterior predictive probabilities.
\newblock {\em Journal of the American Statistical Association \emph{\url{https://doi.org/10.1080/01621459.2026.2721401}}\/}.

\bibitem[\protect\citeauthoryear{Hagar and McGree}{Hagar and McGree}{2026}]{hagar2026economical}
Hagar, L. and J.~M. McGree (2026).
\newblock Computationally efficient experimental design with generalized posteriors.
\newblock {\em arXiv preprint arXiv:2605.00379\/}.

\bibitem[\protect\citeauthoryear{Hagar and Stevens}{Hagar and Stevens}{2023}]{hagar2025precision}
Hagar, L. and N.~T. Stevens (2023).
\newblock An economical approach to design with precision criteria \url{https://arxiv.org/abs/2306.09476}.

\bibitem[\protect\citeauthoryear{Hagar and Stevens}{Hagar and Stevens}{2025}]{hagar2025economical}
Hagar, L. and N.~T. Stevens (2025).
\newblock An economical approach to design posterior analyses.
\newblock {\em Journal of the American Statistical Association\/}~{\em 120\/}(552), 2559--2568.

\bibitem[\protect\citeauthoryear{Hagar and Stevens}{Hagar and Stevens}{2026}]{hagar2024fast}
Hagar, L. and N.~T. Stevens (2026).
\newblock {Fast power curve approximation for posterior analyses}.
\newblock {\em Bayesian Analysis\/}~{\em 21}, 255 -- 280.

\bibitem[\protect\citeauthoryear{Jiang, Wang, Li, Xia, and Jia}{Jiang et~al.}{2012}]{jiang2012practical}
Jiang, Z., L.~Wang, C.~Li, J.~Xia, and H.~Jia (2012).
\newblock A practical simulation method to calculate sample size of group sequential trials for time-to-event data under exponential and {W}eibull distribution.
\newblock {\em PLOS One\/}~{\em 7}, e44013.

\bibitem[\protect\citeauthoryear{Joseph and Belisle}{Joseph and Belisle}{1997}]{joseph1997bayesian}
Joseph, L. and P.~Belisle (1997).
\newblock Bayesian sample size determination for normal means and differences between normal means.
\newblock {\em Journal of the Royal Statistical Society: Series D (The Statistician)\/}~{\em 46\/}(2), 209--226.

\bibitem[\protect\citeauthoryear{Landau and Stahl}{Landau and Stahl}{2013}]{landau2013sample}
Landau, S. and D.~Stahl (2013).
\newblock Sample size and power calculations for medical studies by simulation when closed form expressions are not available.
\newblock {\em Statistical Methods in Medical Research\/}~{\em 22\/}(3), 324--345.

\bibitem[\protect\citeauthoryear{Lindley}{Lindley}{1997}]{lindley1997choice}
Lindley, D.~V. (1997).
\newblock The choice of sample size.
\newblock {\em Journal of the Royal Statistical Society: Series D (The Statistician)\/}~{\em 46\/}(2), 129--138.

\bibitem[\protect\citeauthoryear{Nemirovski, Juditsky, Lan, and Shapiro}{Nemirovski et~al.}{2009}]{nemirovski2009robust}
Nemirovski, A., A.~Juditsky, G.~Lan, and A.~Shapiro (2009).
\newblock Robust stochastic approximation approach to stochastic programming.
\newblock {\em SIAM Journal on Optimization\/}~{\em 19\/}(4), 1574--1609.

\bibitem[\protect\citeauthoryear{Polyak}{Polyak}{1990}]{polyak1990new}
Polyak, B. (1990).
\newblock New stochastic approximation type procedures.
\newblock {\em Avtomatica i Telemekhanika\/}~{\em 7}, 98--107.

\bibitem[\protect\citeauthoryear{Polyak and Juditsky}{Polyak and Juditsky}{1992}]{polyak1992acceleration}
Polyak, B.~T. and A.~B. Juditsky (1992).
\newblock Acceleration of stochastic approximation by averaging.
\newblock {\em SIAM journal on control and optimization\/}~{\em 30\/}(4), 838--855.

\bibitem[\protect\citeauthoryear{Raiffa, Schlaifer, et~al.}{Raiffa et~al.}{1961}]{raiffa1961applied}
Raiffa, H., R.~Schlaifer, et~al. (1961).
\newblock {\em Applied statistical decision theory}.
\newblock Boston: Harvard University Graduate School of Business Administration.

\bibitem[\protect\citeauthoryear{Robbins and Monro}{Robbins and Monro}{1951}]{robbins1951stochastic}
Robbins, H. and S.~Monro (1951).
\newblock A stochastic approximation method.
\newblock {\em The Annals of Mathematical Statistics\/}~{\em 22}, 400--407.

\bibitem[\protect\citeauthoryear{Ruppert}{Ruppert}{1988}]{ruppert1988efficient}
Ruppert, D. (1988).
\newblock Efficient estimators from a slowly convergent robbins-monro procedure.
\newblock {\em School of Oper. Res. and Ind. Eng., Cornell Univ., Ithaca, NY, Tech. Rep\/}~{\em 781}.

\bibitem[\protect\citeauthoryear{Sadatsafavi, Gustafson, Setayeshgar, Wynants, and D~Riley}{Sadatsafavi et~al.}{2026}]{sadatsafavi2026bayesian}
Sadatsafavi, M., P.~Gustafson, S.~Setayeshgar, L.~Wynants, and R.~D~Riley (2026).
\newblock Bayesian sample size calculations for external validation studies of risk prediction models.
\newblock {\em Statistics in Medicine\/}~{\em 45\/}(3-5), e70389.

\bibitem[\protect\citeauthoryear{Shilane, Budugutta, and Bansal}{Shilane et~al.}{2023}]{shilane2023simulation}
Shilane, D., S.~Budugutta, and M.~Bansal (2023).
\newblock {\em nRegression: Simulation-Based Calculations of Sample Size for Linear and Logistic Regression}.
\newblock R package version 0.5.1.

\bibitem[\protect\citeauthoryear{Sommer, Holst, and Shamsi}{Sommer et~al.}{2026}]{sommer2026carts}
Sommer, B., K.~K. Holst, and F.~Shamsi (2026).
\newblock {\em carts: Simulation-Based Assessment of Covariate Adjustment in Randomized Trials}.
\newblock R package version 0.2.0.

\bibitem[\protect\citeauthoryear{Spiegelhalter, Freedman, and Parmar}{Spiegelhalter et~al.}{1994}]{spiegelhalter1994bayesian}
Spiegelhalter, D.~J., L.~S. Freedman, and M.~K. Parmar (1994).
\newblock Bayesian approaches to randomized trials.
\newblock {\em Journal of the Royal Statistical Society: Series A (Statistics in Society)\/}~{\em 157\/}(3), 357--387.

\bibitem[\protect\citeauthoryear{Stevens and Hagar}{Stevens and Hagar}{2022}]{stevens2022cpm}
Stevens, N.~T. and L.~Hagar (2022).
\newblock Comparative probability metrics: Using posterior probabilities to account for practical equivalence in {A}/{B} tests.
\newblock {\em The American Statistician\/}~{\em 76\/}(3), 224--237.

\bibitem[\protect\citeauthoryear{Toulis, Horel, and Airoldi}{Toulis et~al.}{2021}]{toulis2021proximal}
Toulis, P., T.~Horel, and E.~M. Airoldi (2021).
\newblock The proximal robbins--monro method.
\newblock {\em Journal of the Royal Statistical Society Series B: Statistical Methodology\/}~{\em 83\/}(1), 188--212.

\bibitem[\protect\citeauthoryear{Valko, Carpentier, and Munos}{Valko et~al.}{2013}]{valko2013stochastic}
Valko, M., A.~Carpentier, and R.~Munos (2013).
\newblock Stochastic simultaneous optimistic optimization.
\newblock In {\em International Conference on Machine Learning}, pp.\  19--27. PMLR.

\bibitem[\protect\citeauthoryear{van~der Vaart}{van~der Vaart}{1998}]{vaart1998bvm}
van~der Vaart, A.~W. (1998).
\newblock {\em Asymptotic Statistics}.
\newblock Cambridge Series in Statistical and Probabilistic Mathematics. Cambridge University Press.

\bibitem[\protect\citeauthoryear{Wang and Gelfand}{Wang and Gelfand}{2002}]{wang2002simulation}
Wang, F. and A.~E. Gelfand (2002).
\newblock A simulation-based approach to bayesian sample size determination for performance under a given model and for separating models.
\newblock {\em Statistical Science\/}~{\em 17\/}(2), 193--208.

\bibitem[\protect\citeauthoryear{Wilding, Batterham, Calanna, Davies, Van~Gaal, Lingvay, McGowan, Rosenstock, Tran, Wadden, et~al.}{Wilding et~al.}{2021}]{wilding2021once}
Wilding, J.~P., R.~L. Batterham, S.~Calanna, M.~Davies, L.~F. Van~Gaal, I.~Lingvay, B.~M. McGowan, J.~Rosenstock, M.~T. Tran, T.~A. Wadden, et~al. (2021).
\newblock Once-weekly semaglutide in adults with overweight or obesity.
\newblock {\em New England Journal of Medicine\/}~{\em 384\/}(11), 989--1002.

\bibitem[\protect\citeauthoryear{Wilson, Hooper, Brown, Farrin, and Walwyn}{Wilson et~al.}{2021}]{wilson2021efficient}
Wilson, D.~T., R.~Hooper, J.~Brown, A.~J. Farrin, and R.~E. Walwyn (2021).
\newblock Efficient and flexible simulation-based sample size determination for clinical trials with multiple design parameters.
\newblock {\em Statistical methods in medical research\/}~{\em 30\/}(3), 799--815.

\end{thebibliography}

\end{document}